\documentclass[]{article}
\usepackage{amsmath}
\usepackage{amssymb}
\newcommand{\rd}[1]{\mathcal{#1}}

\newcommand{\eqa}{\begin{eqnarray}}
\newcommand{\neqa}{\end{eqnarray}}
\newcommand{\be}{\begin{equation}}
\newcommand{\ee}{\end{equation}}

\def\f{\frac}

\def\la{\langle}

\newcommand{\bra}[1]{\la {#1}|}

\newcommand{\ket}[1]{| #1 \rangle}
\newcommand{\Tr}{\mathrm{Tr}}

\newcommand{\Hil}{\mathcal{H}}
\newcommand{\slc}{SL(2,\mathbb{C})}
\begin{document}
\title{Physical boundary Hilbert space and volume operator\\ in the Lorentzian new spin-foam theory}
\author{You Ding, Carlo Rovelli \\[1mm]
\normalsize \em CPT%
\footnote{Unit\'e mixte de recherche (UMR 6207) du CNRS et des Universit\'es
de Provence (Aix-Marseille I), de la Meditarran\'ee (Aix-Marseille II) et du Sud (Toulon-Var); laboratoire affili\'e \`a la FRUMAM (FR 2291).} , CNRS Case 907, Universit\'e de la M\'editerran\'ee, F-13288 Marseille, EU}
\date{\small\today}
\maketitle\vspace{-7mm}
\begin{abstract}
\noindent A covariant spin-foam formulation of quantum gravity has been recently developed, characterized by a kinematics which appears to match well the one of canonical loop quantum gravity. In this paper we reconsider the implementation of the constraints that defines the model. We define in a simple way the boundary Hilbert space of the theory, considering a slight modification of the embedding of the $SU(2)$ representations into the $SL(2,\mathbb{C})$ ones. We then show directly that all constraints vanish on this space in a weak sense. The vanishing is exact (and not just in the large quantum number limit.) We also generalize the definition of the volume operator in the spinfoam model to the Lorentzian signature, and show that it matches the one of loop quantum gravity, as does in the Euclidean case.
\end{abstract}
%%%%
\section{Introduction}
The spinfoam formalism \cite{Reisenberger:1994aw,spinfoam,spinfoam11,BC1,BC2,spinfoams} and
canonical loop quantum gravity (LQG) \cite{alrev,libro,lqg} can ideally be viewed as the covariant and the canonical versions, respectively, of a background-independent quantum theory of gravity \cite{Rovelli:2010wq}. This scenario is nicely realized in three dimensions \cite{alex}, and there are recent attempts to implement it in quantum cosmology \cite{Ashtekar:2009dn,sfc}. An important step ahead towards the realization of this scenario in the complete four dimensional theory has been taken with the recent introduction of two strictly related spin-foam models whose kinematics appears to match the one of LQG rather well, which we refer to as the new model \cite{epr1,EPR,EPRL,P} and the Freidel-Krasnov-Livine-Speziale (FKLS) model \cite{LS1,FK}. Both of them are motivated by a desire to modify the Barrett-Crane (BC) model \cite{BC2}, based on the vertex amplitude introduced by Barrett and Crane \cite{BC1}. The key problem of the BC model is the fact that \emph{intertwiner} quantum numbers are fully constrained by imposing the simplicity constraints, which are second class, as strong operator equations. But imposing second class constraints strongly may lead to the incorrect elimination of physical degrees of freedom.  It is therefore natural to try to free intertwiner degrees of freedom by imposing the simplicity constraints more weakly: they must be imposed in the quantum theory in such a way that in the classical limit the constraints hold, but all physical degrees of freedom remain free.

Among the several ways proposed to impose these constraints, are the master constraint approach \cite{epr1,EPR,EPRL,P} used for the new model, and the coherent state approach \cite{LS1,FK} used to derive the FKLS model. In addition, the new model \cite{epr1,EPR,EPRL,P} can be obtained also using the coherent state approach \cite{LS2,EP} developed in \cite{LS1,FK}.  In  \cite{euclidean}, we have presented a more straightforward ``matrix element" approach to define the physical boundary Hilbert space of the new model for the Euclidean case.  In this approach, we construct a candidate physical boundary Hilbert space, and then we prove that in this space the matrix elements of all the constraints vanish. In this sense, constraints are imposed \emph{weakly}, rather than strongly as in the BC theory.  The resulting physical boundary state space where the constraints vanish weakly turns out to match that of LQG, and a natural map between the two state spaces can be obtained by identifying eigenstates of the same physical quantities. The fact that the matrix elements of the constraints vanish assures that the constraints hold in the classical limit.  The fact that we obtain the same Hilbert space as the one that is defined by the canonical theory assures us that the space selected is not too small, and all degrees of freedom are free.  Here, we generalize this approach to the Lorentzian signature.

The model we construct contains in fact a slight modification with respect to the one in \cite{epr1,EPR,EPRL,P} (corresponding to a slightly different factor ordering of the constraints). The same modification was already considered by Alexandrov in \cite{sergei}. We show that with the modification the matrix elements vanish \emph{exactly}, and not just in the large quantum number limit, as in previous constructions.

We also derive a volume observable on this space. Since the essential property of the volume operator is that it has contribution only from the nodes of a spin network state, the only possible action of the volume operator is on the intertwiners. That's the reason why there is no generic well-defined volume operator when the intertwiner space is restricted to be one dimensional. In the new model, the way one imposes the simplicity constraints frees intertwiner degrees of freedom, and make it possible to define a non-trivial volume operator. In \cite{euclidean}, the volume operator in the new theory has been derived in the Euclidean case and shown explicitly to match the corresponding LQG canonical operator. In this paper, we generalize the volume observable to the Lorentzian signature, and show that the volumes of the covariant and the canonical theories match.

We work only on a fixed triangulation, and assume that the Barbero-Immirzi parameter $\gamma$ is positive. In section 2, we study the physical boundary space. The volume operator is constructed and shown to match the LQG operator in Section 3.

%%%

\section{Physical boundary Hilbert space of the new theory}

In this section, we use the matrix element approach to construct the physical boundary Hilbert space of the new spin-foam theory with Lorentzian signature \cite{EPRL,P}. We give a brief introduction to the new theory and construct a boundary space, and then we show that this space solves all the constraints \emph{weakly}. We also show this boundary space is isomorphic to that of LQG. At last, we use this boundary space to derive the new amplitude from the BF amplitude.

\subsection{The new model and its boundary space}\label{2.1}
Consider a fixed 4-dimensional triangulation $\Delta$, which is formed by oriented
4-simplices, tetrahedra, triangles, segments and points. The cellular complex $\Delta^*$ dual to this triangulation $\Delta$, is made by faces $f$, edges $e$ and vertices $v$, dual respectively to triangles $f$, tetrahedra $t$ and 4-simplices $v$ of $\Delta$.
The new model is defined on the 2-skeleton of $\Delta^*$,
by a standard spin-foam partition function:
\be
Z=\sum_{j_f,i_e} \prod_f\;(2j_f+1)\;\prod_v\;A_v(j_f,i_e),
\ee
where the sum is over an assignment of an irreducible representation $j_f$ of $SU(2)$
to each face $f$, and over an assignment of an element $i_e$ of a basis in the
space of intertwiners to each edge $e$.
The face amplitude is given by the $SU(2)$ dimension $2j+1$, which is determined in \cite{face} by the structure of the boundary Hilbert space and the condition that amplitudes behave appropriately under compositions. We recall that an
intertwiner is an element of the $SU(2)$ invariant subspace of the
tensor product of the four Hilbert spaces carrying the four
representations associated to the four faces adjacent to a given $e$.
We use the usual basis given by the spin of the virtual link,
under a fixed pairing of the four faces. The amplitude $A_v(j_f,i_e)$ associated to each vertex $v$ is given by
\begin{eqnarray}
A_v(j_{f},i_e)=\sum_{k_e}\int \mathrm{d}p_e (k_e^2+p_e^2)
\left(\prod_{e}\
f^{i_e}_{k_ep_e}(j_{f})\right)
15j_{_{SL(2,\mathbb{C})}}\left((j_{f},\gamma (j_{f}+1));(k_e,p_e)\right),\label{amplitude}
\end{eqnarray}
where the sum and the integral are over an assignment of an irreducible unitary representation $(k,p)$ of $SL(2,\mathbb{C})$ , with $k$ a nonnegtive integer and $p$ real \cite{ruhl,gms}; $15j_{\scriptscriptstyle SL(2,\mathbb{C})}$ is the Wigner 15j symbol of the
group $SL(2,\mathbb{C})$; $f^{i_e}_{k_ep_e}(j_{f})$ is the fusion coefficient obtained contracting $SU(2)$ intertwiners and $SL(2,\mathbb{C})$ intertwiners.
As shown in \cite{epr1,EPR,EPRL,P}, the boundary Hilbert space, satisfying all the kinematical constraints, play a very important role in the construction of the vertex amplitude (\ref{amplitude}). Let us now come to give the boundary Hilbert space.

Given a 3-surface $\Sigma$ intersecting no vertices of $\Delta^*$,
let $\gamma_{\Sigma} := \Delta^* \cap \Sigma$.  We start from the Hilbert space
associated with $\Sigma$ \cite{EPRL,P}:
\begin{equation}
\Hil_{\Sigma} = L^2\left(SL(2,\mathbb{C})^{|L(\gamma_\Sigma)|},\mathrm{d}\mu_{_{\mathrm{Haar}}}\right),
\end{equation}
where $\mu_{\mathrm{Haar}}$ is the Haar measure on the group $SL(2,\mathbb{C})$;
$|L(\gamma_\Sigma)|$ denotes the number of links in
$\gamma_\Sigma$. We fix the orientation such that the node $n = e \cap \Sigma$ is the source of the link $l= f \cap \Sigma$.

By Peter-Weyl theorem, $\Hil_{\Sigma}$ can be decomposed as follows
\begin{equation}
\Hil_{\Sigma} = \bigoplus_{\chi_l}\bigotimes_l \left(\Hil_{\chi_l}^* \otimes \Hil_{\chi_l}\right),
\end{equation}
where $\chi_l$ is an assignment of an $SL(2,\mathbb{C})$ representation to each link $l$ and $\Hil_{\chi}$ is the carrier space of the representation $\chi$. The two Hilbert spaces associated to the link $l$ are naturally associated to the two nodes that bound the link $l$, because they transform under the action of a gauge transformation at one end of the link. Regrouping the four Hilbert spaces associated to each node $n$, the last equation can be rewritten in the form
\begin{equation}
\Hil_{\Sigma} = \bigoplus_{\chi_l}\bigotimes_n \Hil_{n}.
\end{equation}
The Hilbert space associated to a node $n$ is
\begin{equation}
\Hil_{n} = \bigotimes_{a=1}^4 \Hil_{\chi_{a}},
\end{equation}
where $a=1,2,3,4$ runs here over the four links that join at the node $n$ (that is, the four faces of the boundary tetrahedron $t$),
and we have identified the Hilbert space carrying a representation and its dual.
Here the nodes $n$ label the tetrahedra $t$ in the boundary. We restrict our attention to a single boundary tetrahedron, and its associated Hilbert space $\Hil_{n}$, which we call simply $\Hil$ in the following.

Consider the irreducible unitary representations of the principal series of $SL(2,\mathbb{C})$ (for details see \cite{ruhl, gms}), $\Hil:=\Hil_n$ has the structure
\be
\Hil = \bigotimes_{a=1}^{4}\Hil_{(k_a,p_a)},
\ee
with $k$ a nonnegative integer and $p$ real.
The physical intertwiner state space $\mathcal{K}_{\mathrm{ph}}$ is a subspace of this space, where the constraints hold in a suitable sense.

As a first step to give the physical boundary space, let us restrict the representations to the ones that satisfy \cite{sergei}
\be%
p=\gamma (k+1).
\label{gammasimple}
\ee%
We call $\gamma$-simple the $SL(2,\mathbb{C})$ representations that satisfy this relation. With this relation, the continuous label $p$ becomes quantized, because $k$ is discrete. It is
because of this fact that any continuous spectrum depending on $p$ comes out effectively
discrete on the subspace satisfying the relation (\ref{gammasimple}). Notice that the relation here is slightly different from the one used in the literature \cite{EPRL,P}, which is $p=\gamma k$. We will show later that this difference is very important for our construction.

Next, fix an $SU(2)$ subgroup of $SL(2,\mathbb{C})$, then the $(k,p)$ representation for the single component of $\Hil$ associated with a single boundary face $f$ splits into the irreducible representations $\Hil_j$ of the $SU(2)$ subgroup as
\begin{equation}
\mathcal{H}_{(k,p)} = \bigoplus_{j=k}^{\infty}\mathcal{H}_j,\label{CG}
\end{equation}
with $j$ increasing in steps of 1.
Consider the lowest spin term in each factor, where $j$ in the decomposition (\ref{CG}) is reduced to
\begin{equation}
j=k;  \label{jk}
\end{equation}
 this selects the ``minimal" subspace
\begin{equation}
\mathcal{H}^{\mathrm{min}}=\bigotimes_{a=1}^4 \mathcal{H}_{k_a}.\label{min}
\end{equation}
The final physical intertwiner space $\rd{K}_{\rm{ph}}$ is given by the $SU(2)$-invariant subspace of $\mathcal{H}^{\mathrm{min}}$:
\be%
{\cal
K}_{\mathrm{ph}}= \mathrm{Inv}_{\mathrm{SU}(2)}[{\Hil}^{\mathrm{min}}].\label{inv}
\ee%
The total
physical boundary space ${\rd H}_{\mathrm{ph}}$ of the theory is then
obtained as the span of spin-networks in $L^2[SL(2, \mathbb{C})^L/SL(2,\mathbb{C})^N]$ with
$\gamma$-simple representations on edges and with intertwiners in the spaces
${\rd K}_{\mathrm{ph}}$ at each node. In the next subsection, we will show this physical boundary space ${\rd H}_{\mathrm{ph}}$ solves all the kinematic constraints in a suitable sense.
\subsection{Kinematic constraints}
Now let us come to introduce the kinematic constraints, including the simplicity constraints and the closure constraint, and show all of them are satisfied on the physical boundary space ${\rd H}_{\mathrm{ph}}$. On the fixed oriented triangulation $\Delta$, we restrict the metric to be a Regge one \cite{regge}: flat within each 4-simplex, with curvature on the triangles. We choose as the boundary variables $\mathfrak{sl}(2,\mathbb{C})$-valued variables $J_{l}^{IJ}$ associated with the
links $l$ of the graph formed by the one-skeleton of the cellular complex dual to the boundary triangulation. We also need constraints to restrict these variables to the gravity fields, whose discrete forms introduced in \cite{epr1,EPR,EPRL} are given by:
\begin{align}
\rm{simplicity\ constraints:}\quad
&C_l^J=n_I\left((^*
J_l)^{IJ}+\frac{1}{\gamma}J_l^{IJ}\right) = 0 \label{simplicity},\\
\rm{closure\ constraints:}\quad&G^{IJ}=\sum_{a=1}^4 J^{IJ}_{l_a} = 0,\label{closure}
\end{align}
where $n_I$ denotes the normal to the tetrahedron $t$, and $*$ stands for the Hodge dual in the internal indices, the completely antisymmetric objects $\epsilon^{IJKL}$ defined as $\epsilon^{0123}=1$ ; the sum is over the four links that join at the node dual to the tetrahedron (or over the four triangles bound the tetrahedron).
These constraints will give the solution $B_f=\int_{f}{}^*\big(e(t)\wedge e(t)\big)$, where $e(t)$ is a tetrad one-form covering the tetrahedron $t$, and $J_f=B_f+\frac{1}{\gamma}{}^*B_f$, with triangle $f$ dual to the link $l$.
The usual
\begin{align}
\mathrm{quadratic\ diagonal}\quad
&C_{ll}:=\Big(1-\frac{1}{\gamma^2}\Big){}^*J_l\cdot J_{l}+\frac{2}{\gamma}J_l\cdot J_{l}= 0
\label{c1}\\
\mathrm{and\ off-diagonal}\quad
&C_{ll'}:=\Big(1-\frac{1}{\gamma^2}\Big){}^*J_l\cdot J_{l'}+\frac{2}{\gamma}J_l\cdot J_{l'}= 0
\label{c0}
\end{align}
simplicity constraints can be easily shown to follow from (\ref{simplicity}).
This reformulation is central for the new models \cite{epr1,EPR,EPRL,P,LS1,FK,LS2}.
In particular, if we choose a ``time" gauge where $n_I=(0,0,0,1)$, the
simplicity constraint (\ref{simplicity}) turns out to be
\begin{eqnarray}
C_l^i=J^{0i}_l+\gamma\,{}^*J^{0i}_l=0\label{gaugefixsimplicity},
\end{eqnarray}
which leads to the key constraint of the new model
\be
C_l^i=K^i_l+\gamma\ L^i_l=0,
\label{key}
\ee
where $L_l^j:=\frac{1}{2}\epsilon^{j}{}_{kl}J_l^{kl}$ and $K_l^j:=J_l^{0j}$ are respectively the generators of the $SU(2)$ subgroup that leaves $n_I$ invariant, and the generators of the corresponding boosts. In terms of these generators, the closure constraint (\ref{closure}) becomes
\begin{subequations}\label{closureKL}
\begin{align}
G_{L}^i&=\sum_{a=1}^4L^i_l=0\label{closureL}\\
\mathrm{and}\quad G_{K}^i&=\sum_{a=1}^4K^i_l=0 \label{closureK}.
\end{align}
\end{subequations}

To quantize the constraints (\ref{key}) (\ref{closureKL}), one just need to replace the generators with the associated operators.
Given a carrier space $\Hil_{(k,p)}$, the canonical basis is given by the basis diagonalizing simultaneously the Casimir operators $J\cdot J, {}^*J\cdot J, L\cdot L$ and $L^3$ , which is noted as $\ket{(k,p);j,m}$ or simply as $\ket{j,m}$. On this canonical basis, the generators act in the
following way \cite{gms}:
\begin{align}
L^3\ket{j,m}=& m\ket{j,m}, \nonumber \\
L^+\ket{j,m}=& \sqrt{(j+m+1)(j-m)}\ket{j,m+1}, \nonumber\\
L^-\ket{j,m}=& \sqrt{(j+m)(j-m+1)}\ket{j,m-1}, \nonumber \\
K^3\ket{j,m}=& -\alpha_{(j)}\sqrt{j^2-m^2}\ket{j-1,m}-\beta_{(j)}m\ket{j,m}
+\alpha_{(j+1)}\sqrt{(j+1)^2-m^2}\ket{j+1,m}, \nonumber \\
K^+\ket{j,m}=& -
\alpha_{(j)}\sqrt{(j-m)(j-m-1)}\ket{j-1,m+1}-\beta_{(j)}\sqrt{(j-m)(j+m+1)}
\ket{j,m+1} \label{lorentzrep}\\
&-\alpha_{(j+1)}\sqrt{(j+m+1)(j+m+2)}\ket{j+1,m+1},\nonumber \\
K^-\ket{j,m}=&
\alpha_{(j)}\sqrt{(j+m)(j+m-1)}\ket{j-1,m-1}-\beta_{(j)}\sqrt{(j+m)(j-m+1)}
\ket{j,m-1} \nonumber \\
&+\alpha_{(j+1)}\sqrt{(j-m+1)(j-m+2)}\ket{j+1,m-1}, \nonumber
\end{align}
where
\begin{align}
& L^{\pm}=L^1\pm iL^2,\qquad K^{\pm}=K^1\pm iK^2\nonumber\\
\mathrm{and} \quad &\alpha_{(j)}=\frac{i}{j}\sqrt{\frac{(j^2-k^2)(j^2+p^2)}{4j^2-1}}, \qquad\beta_{(j)}=\frac{kp}{j(j+1)}
.
\end{align}

Now let us go to show the physical Hilbert space ${\rd H}_{\mathrm{ph}}$ derived last subsection solves indeed the constraint operators associated to the simplicity constraints (\ref{key}) and the closure constraints (\ref{closureKL}). Namely, we will show
\begin{itemize}
\item[(i)] the simplicity constraints (\ref{key}) are satisfied in the ``minimal'' $\gamma$-simple representation $\Hil^{\rm{min}}$,
\item[(ii)] the closure constraints (\ref{closureKL}) are satisfied in the intertwiner space $\mathcal{K}_{\mathrm{ph}}$.
\end{itemize}

To show (i), let us consider the states in the ``minimal'' space $\Hil^{\rm{}min}$ in equation (\ref{min}). For these lowest spin states, equation \eqref{jk} implies that the states are of the form $\ket{(k,p);k,m}$, or simply as $\ket{k,m}$. The action (\ref{lorentzrep}) of the generators on these states reads:
\begin{align}
\big(K^{3}+\beta_{(k)}L^{3}\big)\ket{k,m}&=\alpha_{(k+1)}\sqrt{(k+1)^2-m^2}\ket{(k+1,m)},\nonumber\\
\big(K^{+}+\beta_{(k)}L^{+}\big)\ket{k,m}&=-\alpha_{(k+1)}\sqrt{(k+m+1)(k+m+2)}\ket{(k+1,m+1)},\nonumber\\
\big(K^{-}+\beta_{(k)}L^{-}\big)\ket{k,m}&=\alpha_{(k+1)}\sqrt{(k-m+1)(k-m+2)}\ket{(k+1,m-1)}.\nonumber
\end{align}
It is straightforward to obtain
\begin{align}
\bra{k,m'}\big(K^i+\beta_{(k)} L^i\big)\ket{k,m}=0.\label{beta}
\end{align}
Using the relation (\ref{gammasimple}), $\beta_{(k)}$ turns out to be the Barbero-Immirzi parameter $\gamma$ and the matrix elements of the l.h.s of (\ref{key}) hence vanish on the ``minimal'' $\gamma$-simple space:
\begin{align}
\bra{k,m'}C^i\ket{k,m}=\bra{k,m'}\big(K^i+\gamma L^i\big)\ket{k,m}=0.\label{keyme}
\end{align}
Notice that the slight difference of our relation (\ref{key}) from the old one plays a key role here.  Notice also that what we obtain is that the matrix elements vanish \emph{exactly}, and not just in the large spin limit.

To show (ii), observe that the l.h.s. of (\ref{closureL}) is the generator of $SU(2)$ transformations at the node and vanishes strongly on (\ref{inv}) by definition; the l.h.s. of (\ref{closureK}) is proportional to the one of (\ref{closureL}) by (\ref{keyme}) and therefore vanishes weakly.
Thus $\mathcal{K}_{\mathrm{ph}}$ is the intertwiner space as a solution of \emph{all} the constraints: all the constraints hold weakly.

Notice that the intertwiner space $\mathcal{K}_{\mathrm{ph}}$ is not $\slc$-invariant, but only $SU(2)$-invariant, since we impose the closure constraint weakly, instead of  strongly. One can  impose the closure constraint strongly to get an $\slc$-invariant Hilbert space, as in \cite{projected}, but we still prefer this construction, since the resulting space is naturally isometric to the LQG one, while its projection on the $\slc$-invariant states is not  \cite{Kaminski:2009cc}. (Canonical quantization in a fixed gauge, as the one used in LQG, is generally reliable for determining the correct Hilber space.)  As we shall see in the next section, Lorentz invariance is fully implemented by the transition amplitudes. 

Summarizing, we have introduced the kinematic constraints and shown that all of them are satisfied on the physical boundary space $\Hil_{\mathrm{ph}}$ derived in the last subsection. Since we have not proven that the physical Hilbert space considered is the \emph{maximal} space where the constraints hold weakly, one might worry that the physically correct quantization of the degrees of freedom of general relativity could need a larger space. Also, it has been pointed out that imposing second class constraints weakly might lead to inconsistencies in some cases \cite{sergei}. In the present case, however, these worries are not relevant, since the space obtained is directly related to the one of the canonical theory, which we can trust to capture the degrees of freedom of gravity correctly.

\subsection{Dynamics}

We have the remarkable result that $\mathcal{K}_{\mathrm{ph}}$ is naturally isomorphic to the $SU(2)$ intertwiner space, and therefore the
constrained boundary space ${\rd H}_{\mathrm{ph}}$ can be identified with the $SU(2)$
LQG state space ${\rd H}_{{SU}(2)}$ associated to the graph which is
dual to the boundary of the triangulation, namely the space of the
$SU(2)$ spin networks on this graph.
For completeness, let us repeat some materials in \cite{EPRL, P} to exhibit this isomorphism by the embedding of the Hilbert space of LQG $\Hil_{SU(2)}$ into the boundary Hilbert space $\Hil_{\mathrm{ph}}$ of the new model;  we also use this embedding and the BF amplitude of a single 4-simplex $v$ to derive the amplitude  (\ref{amplitude}).

The way we construct the boundary space gives a projection, which maps simple $SL(2,\mathbb{C})$ spin-network states to $SU(2)$ spin-network states. The corresponding embedding $f$ is defined as the hermitian conjugate of this projection, which is given by
\begin{align}
f_{(j)}:\quad\quad\Hil_{j}&\longrightarrow \Hil_{( j,\gamma(j+1))},\nonumber\\
\ket{j,m}&\longmapsto \ket{( j,\gamma(j+1));j,m}
\end{align}
for the representations and
\begin{align}
f_{(j_l)}:\ &\mathrm{Inv}_{_{SU(2)}}(\otimes_{a=1}^4 \Hil_{j_a})\longrightarrow \mathrm{Inv}_{_{SL(2,\mathbb{C})}}(\otimes_{a=1}^4 \Hil_{( j_a,\gamma(j_a+1))}),\nonumber\\
&{i}^{m_1m_2m_3m_4}\longmapsto \int\limits_{SL(2,\mathbb{C})}\mathrm{d}g\, \Big(\prod_{a=1}^4D^{( j_a,\gamma(j_a+1))}(g)^{(j'_a,m'_a)}{}_{(j_a,m_a)}\Big)i^{m_1m_2m_3m_4},
\end{align}
for intertwiners \cite{EPRL, P},
where $D^{(k,p)}(g)^{(j',m')}{}_{(j,m)}$ denote the matrix elements of the irreducible representation $(k,p)$, with indices $(j,m)$.
Let indices $(j,m)\equiv \alpha$, ${\chi}^{\alpha_1\alpha_2\alpha_3\alpha_4}$ denote the $SL(2,\mathbb{C})$ intertwiner defined by a virtual link carring the representation $\chi=(k,p)$, and $\mathrm{d}\chi$ the Plancherel measure on the spectrum. Then using the relation
\begin{align}
\int\limits_{SL(2,\mathbb{C})}\mathrm{d}g\,D^{(\chi_1)}(g)^{\alpha_1}{}_{\alpha'_1}D^{(\chi_2)}(g)^{\alpha_2}{}_{\alpha'_2}\overline{D^{(\chi_3)}}(g)^{\alpha_3}{}_{\alpha'_3}\,\overline{D^{(\chi_4)}}(g)^{\alpha_4}{}_{\alpha'_4}
=\int\mathrm{d}\chi\, {\chi}^{\alpha_1\alpha_2\alpha_3\alpha_4}\overline{{\chi_{\alpha'_1\alpha'_2\alpha'_3\alpha'_4}}},
\end{align}
one can obtain
\begin{align}
f_{(j_l)}\ket{i}=\int \mathrm{d}\chi\, f^i_{\chi}(j_l) \ket{{\chi}},
\end{align}
where the coefficients $f^i_{\chi}(j_l)$ are given by
\begin{align}
f^i_{\chi}(j_l)=i^{m_1m_2m_3m_4}\overline{{\chi}}{}_{(j_1,m_1)(j_2,m_2)(j_3,m_3)(j_4,m_4)}.
\end{align}
If we piece these maps at each node, we obtain the map $f: \Hil_{SU(2)}\rightarrow\Hil_{\mathrm{ph}}$ of the entire LQG space into the state space of the new theory. In the spin network basis we obtain
\begin{align}
f_{(j_l)}:\ket{j_l,i_n}\mapsto\int\mathrm{d}\chi_n\,f^{i_n}_{\chi_n}(j_l)\ket{(j_l,\gamma(j_l+1)),\chi_n}\label{emb}.
\end{align}

Now let us use this embedding and the BF amplitude to give the new amplitude (\ref{amplitude}).
The BF amplitude of a single 4-simplex $v$ for a given boundary state $\ket{\Psi}$ reads
\begin{align}
A(\Psi)=\int\limits_{SL(2,\mathbb{C})^{10}}\prod_{f}\mathrm{d}g_f\ \Psi(g_f)\int\limits_{SL(2,\mathbb{C})^{5}}\prod_{e}\mathrm{d}V_e\prod_f\delta(V_{e_f}\,g_f\,V^{-1}_{e'_f}),
\end{align}
where  $V_{e_f}, V_{e'_f}$ are the two group elements around the perimeter of $f$, which is in the 4-simplex $v$ and not in the boundary. The integral over $g_f$ gives
\begin{align}
A(\Psi)=\int\limits_{SL(2,\mathbb{C})^{5}}\prod_{e}\mathrm{d}V_e\ \Psi(V_{e_f}^{-1}V_{e'_f}).
\end{align}
In the new theory, for any boundary state $\Psi\in\Hil_{\mathrm{ph}}$,  according to the embedding (\ref{emb}), there exist a LQG state $\psi\in\Hil_{\mathrm{LQG}}$, such that $\Psi=f(\psi)$. Let us consider the specific case when $\psi$ is a spin-network state $\ket{\psi}=\ket{j_f,i_e}$ on the boundary. The amplitude is then given explicitly by
\begin{align}
A(j_f,i_e)=\int\mathrm{d}\chi_e\big(\prod_ef^{i_e}_{\chi_e}(j_f)\big)15j((j_f,\gamma(j_f+1)),\chi_e)\label{amplitud}.
\end{align}
In terms of $(k,p)$, the Plancherel measure $\mathrm{d}\chi$ can be expressed as $(k^2+p^2)\mathrm{d}p$, which gives the expression (\ref{amplitude}).

\section{Geometrical observables}
The \emph{kinematics} of canonical loop quantum gravity is rather well understood; in particular, the properties of the geometrical operators, including the area and the volume operators\cite{RS,area,volume} are well established. (On the volume operator, see also \cite{semi}.)
In this section, we study the geometrical operators in the new spinfoam model and their relation with the $SU(2)$ ones in LQG.
\subsection{The area operator}
The area operator of the new spinfoam model has been derived in \cite{EPRL,EP} and shown to match the LQG one.
Classically, the area $A(f)$ of a triangle $f$ is given by $A(f)^2={\frac{1}{2}({}^*B_f)^{IJ}\cdot({}^*B_f)_{IJ}}$. If we fix the time gauge, we have
$A_3(f)^2={\frac{1}{2}({}^*B_f)^{ij}\cdot({}^*B_f)_{ij}}$. These two quantities are equal up to a constrained term. As shown in \cite{EP,EPRL}, using the constraints, the operator related to $A_{3}(f)^2$ can be obtained as $A_3(f)^2=\kappa^2\gamma^2L_f^2$, which matches three-dimensional area as determined by LQG, including for the correct Barbero-Immirzi parameter proportionality factor.
\subsection{The volume operator}
Let us now turn to study the volume operator.
Following \cite{euclidean}, the volume of a boundary tetrahedron $t$ is $V(t)=\sqrt{|V^2(t)|}$ where 
\be%
V^2(t)={\f{1}{27}\epsilon^{abc}\Tr[{}^*B_a{}^*B_b{}^*B_c]},
\ee%
which in terms of $J$ turns out to be
\begin{align}
&V^2(t)={\f{1}{27}\Big(\frac{\gamma^2}{1+\gamma^2}\Big)^3\epsilon^{abc}\Tr\Big[\Big(\frac{1}{\gamma}J_a+{}^*J_a\Big)\Big(\frac{1}{\gamma}J_b+{}^*J_b\Big)\Big(\frac{1}{\gamma}J_c+{}^*J_c\Big)\Big]}
\label{v4j}
\end{align}
The volume operator $\hat{V}(t)$ of the tetrahedron $t$ is
then formally given by (\ref{v4j}) with $J^{IJ}$ replaced by the corresponding operators:
\begin{align}
\widehat{V^2}(t)={\f{1}{27}\Big(\frac{\gamma^2}{1+\gamma^2}\Big)^3\epsilon^{abc}\Tr\Big[\Big(\frac{1}{\gamma}\hat{J}_a+{}^*\hat{J}_a\Big)\Big(\frac{1}{\gamma}\hat{J}_b+{}^*\hat{J}_b\Big)\Big(\frac{1}{\gamma}\hat{J}_c+{}^*\hat{J}_c\Big)\Big]}.
\label{hatv}
\end{align}
Since the volume operator does not change the graph of the spin network sates, nor the coloring of the links, its action can be studied on the Hilbert space associated to a single node. Consider the matrix element of the square of the volume operator between two states in the physical intertwiner space (we drop the hats):
\be
\bra{i} \widehat{V^2} \ket{j}
=\f{1}{27}\big(\f{\gamma^2}{1+\gamma^2}\big)^3\epsilon^{abc}\bra{i} \big(\f{1}{\gamma}J_a^{ij}+{}^*J^{ij}_a\big)\big(\f{1}{\gamma}J_b^{jk}+{}^*J^{jk}_b\big)\big(\f{1}{\gamma}J_c^{ki}+{}^*J_c^{ki}\big) \ket{j}.
\ee
Writing this in terms of $L$ and $K$ components gives
\be
\bra{i} \widehat{V^2} \ket{j}
=\f{1}{27}\big(\f{\gamma^2}{1+\gamma^2}\big)^3\epsilon^{abc}\epsilon^{ij}_{\ \ m}\epsilon^{jk}_{\ \ n}\epsilon^{ki}_{\ \ p}\bra{i} \big(\f{1}{\gamma}L_a^{m}-K^{m}_a\big)\big(\f{1}{\gamma}L_b^{n}-K^{n}_b\big)\big(\f{1}{\gamma}L_c^{p}-K_c^{p}\big) \ket{j}.\label{simplify}
\ee
Notice that the intertwiner space is the subspace of the product of the space $\Hil_{a}$ associated to the link $a$, and the action of $(K_a,\,L_a)$ is in fact on $\Hil_a$. Hence we can use the form  (\ref{key}) of the simplicity constraint to simplify Eq.~(\ref{simplify}), although the r.h.s seems a polynomial.
Using the form (\ref{key}) of the constraint, we can rewrite it as
\begin{align}
\bra{i} \widehat{V^2} \ket{j}
=&\f{1}{27}\big(\f{\gamma^2}{1+\gamma^2}\big)^3\big(\f{1}{\gamma}+\gamma\big)^3\epsilon^{abc}\epsilon_{ijk}\bra{i} L_a^{i}L_b^{j}L_c^{k} \ket{j}
\end{align}
and a little algebra gives
\be
\bra{i} \widehat{V^2} \ket{j}
=\Big(\frac{\gamma}{3}\Big)^3 \ \epsilon^{abc}\epsilon_{ijk}\bra{i} L_a^{i}L_b^{j}L_c^{k} \ket{j}.
\ee
That is
\be
\hat V(t)
={\Big(\frac{\gamma}{3}\Big)}^{\frac32} \sqrt{\left|\epsilon^{abc}\epsilon_{ijk} L_a^{i}L_b^{j}L_c^{k}\right|}.
\ee
Now, the operator on the r.h.s. is precisely the LQG volume operator $V(t)_{\mathrm{LQC}}$ of the tetrahedron, as it acts on ${\cal
K}_{\mathrm{ph}}$ including the correct dependence on the Barbero-Immirzi parameter $\gamma$.
\section*{Acknowledgments}
Thanks to Eugenio Bianchi for enlightening discussions. We wish also to thank Jonathan Engle, Simone Speziale, Franck Hellmann, Roberto Pereira, Muxin Han and Sergei Alexandrov for helpful comments and suggestions. Y. D. is supported by CSC scholarship No. 2008604080.

\end{document}